\documentstyle[preprint,aps]{revtex}

\begin{document}
\title{Self-Consistent Born Approximation for the Hole
Motion in the Three-Band Model: a Comparison with Photoemission Experiments}
\author{Oleg A. Starykh~\cite{a,oas}, Oziel F. de Alcantara Bonfim,
George F. Reiter}
\address{Texas Center for Superconductivity and Physics Department,
University of Houston, Houston, TX 77204-5932}
\date{June 14, 1995}
\maketitle{}
\begin{abstract}
The dispersion relation of the single hole in $CuO_2$ plane is
calculated in the self-consistent Born approximation for the
thee-band Hamiltonian. We find that direct oxygen-oxygen
hopping removes the strong anisotropy of the hole spectrum around the
band minima.
Our results compare well with recent photoemission measurements
of single hole dispersion relation in the $Sr_2 CuO_2 Cl_2$.

\end{abstract}

Recent angle resolved photoemission measurements on different
copper-oxide materials \cite{photo} have found a number of interesting
features in the quasiparticle dispersion. The most interesting is the
finding of an almost dispersionless (flat) regions around $(\pi,0)$ and
$(0,\pi)$ points of the Brilloin zone \cite{flat}. These flat bands
were observed in doped Bi2212 and YBaCuO compounds. Recently, APRES
experiment was performed on $Sr_2 CuO_2Cl_2$ \cite{wells}, which
is an $insulating$ layered cooper oxide. The experiment has the virtue
of measuring a $single$ hole dispersion relation in a quantum antiferromagnet.
It was found that the overall bandwidth ($\sim 2.2 J$) and the position
of the minimum of the hole band at $(\pi/2,\pi/2)$ agree with the
$t-J$ model predictions\cite{manous}. These observations leave no doubts that
we are dealing with a hole coupled strongly to the
antiferromagnetically interacting localized spins.

 At the same time some details of the experimentally found spectrum
disagree with these of the $t-J$ model. First of all, dispersion along
the magnetic Brilloin zone boundary ($(0,\pi)\rightarrow (\pi,0)$ line)
is as strong as along $(0,0) \rightarrow (\pi,\pi)$ line. Second, it
turns out that dispersion along $(0,0) \rightarrow (\pi,0)$ line
is very weak, i.e. the band is almost flat, and the hole's energies at $(0,0)$
and $(\pi,0)$ points were found to coincide within the experimental
error \cite{wells}.

 To understand these experimental results we study the motion of a single
hole in the three-band model of $CuO_2$ plane. The Hamiltonian of the
model is \cite{emery}
\begin{eqnarray}
H=&&J\sum_{m,g} \vec{S}_{m} ~\vec{S}_{m+2g} + 2(t_a + t_b)
\sum_{(m,g,g')} \sum_{\sigma,\sigma'} \vec{S}_{m}~a^{+}_{m,g,\sigma}
\vec{s}_{\sigma,\sigma'} a_{m,g',\sigma'}\nonumber\\
&&+\frac{t_b - t_a}{2} \sum_{(m,g,g')} \sum_{\sigma} ~a^{+}_{m,g,\sigma}
a_{m,g',\sigma} ~-|t_{pp}|\sum_{(m,m',g,g')} \sum_{\sigma}
(a^{+}_{m,g,\sigma}~a_{m',g',\sigma} + H.c.)
\end{eqnarray}
Fermion operators $a$ are defined on the oxygen sites,
$\vec{s}_{\sigma,\sigma'} =\frac{1}{2}\vec{\sigma}_{\sigma,\sigma'}$,
$\vec{S}_m$ is a
localized copper $S=1/2$ spin, $\sum_{(m,g,g')}$ represents independent sum
over the four oxygen sites surrounding cooper site $m$, while
$\sum_{(m,m',g,g')}$ means sum over nearest neighbors on the
$oxygen$ lattice. Effective hoppings $t_a$ and $t_b$ are due to the
hybridization $t_{pd}$ between O and Cu orbitals in the original
model \cite{emery}, whereas $t_{pp}$ is an independent microscopic parameter.

 With the help of the canonical transformation
\begin{eqnarray}
&&b_{k,\sigma} = \frac{1}{\rho_k}\left((1 + e^{ik_x})~a_{k,x,\sigma}
+ (1 + e^{ik_y})~a_{k,y,\sigma}\right)\nonumber\\
&&d_{k,\sigma}=\frac{1}{\rho_k}\left(e^{ik_x /2} \cos{k_y/2} ~a_{k,x,\sigma}
 - e^{ik_y /2} \cos{k_x/2} ~a_{k,y,\sigma}\right) \\
&&a_{k,x,\sigma} =\frac{1}{\sqrt{N}} \sum_m a_{m,g_x,\sigma}
e^{i\vec{k}\vec{r}_m},
{}~~\rho_k = 2\sqrt{1 + \gamma_k},
{}~~\gamma_k= \frac{1}{2}(\cos{k_x} + \cos{k_y})\nonumber
\end{eqnarray}
The hole's part of the Hamiltonian may be rewritten as
\begin{eqnarray}
H=&&\frac{2(t_a + t_b)}{N} \sum_{k,k',\sigma,\sigma'} \rho_k \rho_{k'}
b^{+}_{k,\sigma'} b_{k',\sigma} \vec{s}_{\sigma,\sigma'} \vec{S}_{k-k'}
 + \frac{t_b - t_a}{2} \sum_{k,\sigma} \rho_k^2 ~b^{+}_{k,\sigma} b_{k,\sigma}
\\
%&& -4|t_{pp}| \sum_{k,\sigma} \big( \frac{(1 + \cos{k_x})(1 + \cos{k_y})}
%{\rho_k^2} (b^{+}_{k,\sigma} b_{k,\sigma} - d^{+}_{k,\sigma} d_{k,\sigma}) \\
%&&+ \frac{(\cos{k_y} - \cos{k_x})\cos{k_x/2} \cos{k_y/2}}{\rho_k^2}
%(d^{+}_{k,\sigma} b_{k,\sigma} + b^{+}_{k,\sigma} d_{k,\sigma})\big)
&& -4|t_{pp}| \sum_{k,\sigma} \left(
f_1(k)(b^{+}_{k,\sigma} b_{k,\sigma} - d^{+}_{k,\sigma} d_{k,\sigma}) +
f_2(k)(d^{+}_{k,\sigma} b_{k,\sigma} + b^{+}_{k,\sigma} d_{k,\sigma})
\right), \nonumber
\end{eqnarray}
where we introduced short-hand notations for
\begin{eqnarray*}
f_1(k)= &&\frac{(1 + \cos{k_x})(1 + \cos{k_y})}{\rho_k^2},\nonumber\\
f_2(k)=&&\frac{(\cos{k_y} -\cos{k_x})\cos{k_x/2} \cos{k_y/2}}{\rho_k^2}
 \nonumber
\end{eqnarray*}
 Note that spin-fermion interaction term couples $Cu$ spins only to
the symmetric
combination of oxygen fermion operators, i.e. to the $b$-operators.
Another, $d$-band, appears only in direct $O-O$ hopping.

 We will treat the copper spins within the spin wave approximation
\cite{prelov}
\begin{eqnarray}
S^{\pm}_m = &&S(1 \pm e^{i\vec{q}_0\vec{r}_m})\alpha_m +
S(1\mp e^{i\vec{q}_0\vec{r}_m})\alpha^{+}_m,\\
S^{z}_m = &&e^{i\vec{q}_0\vec{r}_m} (S - \alpha^{+}_m \alpha_m)\nonumber
\end{eqnarray}
%S^{+}_q = &&S(\alpha_q + \alpha_{q + q_0} + \alpha^{+}_{-q} -
%\alpha^{+}_{-q-q_0}),\nonumber\\
%S^{-}_q = &&S(\alpha_q - \alpha_{q + q_0}+ \alpha^{+}_{-q} +
%\alpha^{+}_{-q-q_0}),\\
%S^z_q = &&S - \alpha^{+}_q \alpha_q \nonumber
%\end{eqnarray}
 Magnon operators are defined in the whole original Brilloin zone,
and $q_0 \equiv (\pi,\pi)$. The spin Hamiltonian is diagonalized then by
the standard Bogoliubov transformation
\begin{eqnarray}
&&\left(\begin{array} {c} \alpha_{-q} \\ \alpha^{+}_q \end{array}
\right) = \left( \begin{array} {c} u_q ~~v_q\\ v_q ~~u_q \end{array} \right)
\left( \begin{array} {c} \beta_{-q} \\ \beta^{+}_q \end{array} \right), \\
&&u_q= \sqrt{\frac{1 + \omega_q}{2\omega_q}},~~v_q= -sgn(\gamma_q)
\sqrt{\frac{1 - \omega_q}{2\omega_q}}, ~~\omega_q = 4JS \sqrt{1 - \gamma_q^2}
\nonumber
\end{eqnarray}
 It is convenient, following \cite{prelov}, to introduce staggered
fermion operators as well
\begin{eqnarray}
&&b^{+}_{k,\sigma}= \frac{1}{2} \sum_s \left(c^{+}_{k,s} + 4\sigma s
{}~c^{+}_{k+q_0,s} \right)\nonumber\\
&&d^{+}_{k,\sigma}= \frac{1}{2} \sum_s \left( z^{+}_{k,s}
+4\sigma s ~z^{+}_{k+q_0,s} \right),~~s,\sigma =\pm \frac{1}{2}
\end{eqnarray}
 After somewhat lengthly algebra we find that $H=H_0 + H_{int}$, where
the free part is
\begin{eqnarray}
H_0= && \sum_q \omega_q \beta^{+}_q \beta_q + \sum_{k,\sigma,\sigma'}
\epsilon_k \sigma^z_{\sigma,\sigma'} ~c^{+}_{k,\sigma} c_{k,\sigma'}
\nonumber\\
&&+2(t_b - t_a)\sum_{k,\sigma}\left(c^{+}_{k,\sigma} c_{k,\sigma} +
\gamma_k c^{+}_{k,\sigma} c_{k,-\sigma} \right) + H_{pp}
\end{eqnarray}
Here $\epsilon_k = \frac{t_a + t_b}{4} \rho_k \rho_{k+q_0}$ is a free
dispersion due to the $\sigma^z S^z$ coupling in the Hamiltonian (3).
Note that free dispersion term is absent in the $t-J$ model.
In the new basis direct oxygen-oxygen hopping looks a bit
complicated
\begin{eqnarray}
H_{pp}= &&-4|t_{pp}| \sum_k \big( f_1(k) \sum_{\sigma= +,-}(c^{+}_{k,\sigma}
c_{k,\sigma} - z^{+}_{k,\sigma} z_{k,\sigma}) \nonumber\\
&&+ f_1(k+q_0)\left(
(c^{+}_{k,+} - c^{+}_{k,-})(c_{k,+} - c_{k,-}) - (z^{+}_{k,+} - z^{+}_{k,-})
(z_{k,+} - z_{k,-})\right) \\
&&+\left\{f_2(k) (z^{+}_{k,+} + z^{+}_{k,-})(c_{k,+} + c_{k,-})
+ f_2(k+q_0) (z^{+}_{k,+} - z^{+}_{k,-})(c_{k,+} - c_{k,-}) + H.c.\right\}
\big), \nonumber
\end{eqnarray}
where $+~(-)$ subscript denotes $+\frac{1}{2}~(-\frac{1}{2})$ spin projection.

 The advantage of this representation is that interaction part of the
Hamiltonian has a simple form
\begin{equation}
H_{int}=\frac{1}{\sqrt{N}} \sum_{k,q,\sigma,\sigma'} U_{\sigma,\sigma'}(k,q)
\left( c^{+}_{k-q,\sigma} c_{k,\sigma'} \beta^{+}_q + c^{+}_{k,\sigma'}
c_{k-q,\sigma} \beta_q \right)
\end{equation}
For completeness we give an explicit form of the interaction vertex U
here
\begin{eqnarray}
U(k,q)= &&
\frac{t_a + t_b}{4}
\left( \begin{array} {c}
u_q\rho_{k,-}\rho_{k-q,+} + v_q\rho_{k,+}\rho_{k-q,-}
{}~~~u_q\rho_{k,+}\rho_{k-q,+} + v_q\rho_{k,-}\rho_{k-q,-} \\
u_q\rho_{k,-}\rho_{k-q,-} + v_q\rho_{k,+}\rho_{k-q,+}
{}~~~u_q\rho_{k,+}\rho_{k-q,-} + v_q\rho_{k,-}\rho_{k-q,+}
\end{array} \right) , \\
\nonumber\\
&&\rho_{p,-} \equiv \rho_p - \rho_{p+q_0},~~\rho_{p,+} \equiv \rho_p +
\rho_{p+q_0}.\nonumber
\end{eqnarray}

 To calculate the hole's Green's function
\begin{eqnarray}
G(k,\omega)=&&[G_0^{-1}(k,\omega) - \Sigma(k,\omega)]^{-1},\\
G_0(k,\omega) =&& \left(\omega - H_0 \right)^{-1} \nonumber
\end{eqnarray}
we use the self-consistent Born approximation \cite{klr}
\begin{equation}
\Sigma_{\sigma,\sigma'}(k,\omega)= \frac{1}{N}\sum_{\nu,\nu'}
\sum_q U_{\nu,\sigma}
(k,q) G_{\nu,\nu'} (k - q,\omega -\omega_q) U_{\nu',\sigma'} (k,q)
\label{A}
\end{equation}
 In the presence of nonzero oxygen-oxygen hopping $t_{pp}$,
$G(k,\omega)$ is $4\times 4$ matrix,
but the interaction vertices $U_{\nu,\sigma}(k,q)$ are $2\times 2$ matrices,
because there is no coupling between magnons and $z$-fermions.

 The self-consistent Born approximation (SCBA) consists in neglecting
vertex corrections to the self-energy. It was found to work well for
the $t-J$ model \cite{manous}, and was applied to the three-band
model in Refs.\cite{oz,kaban}. It predicts the band minima at $(\pi/2,\pi/2)$
and bandwidth proportional to $J$ \cite{oz}, as does the $t-J$ model.
The energy dispersion, in the absence of $t_{pp}$, is very similar in
the two models.
The main difference is that hole has a spin 1/2. For momenta close to
$(\pi/2,\pi/2)$ the hole's spin is antialigned with the Copper spin, thus
forming a singletlike excitation \cite{oz}. However, away from
minima a strong mixture of the triplet state appears \cite{oz,fren}.

 Omitting $H_{int}$ for a moment we see that at $t_{pp}=0$ the spectrum
has a degenerate minima along the $\gamma_q=0$ curve. Turning on $t_{pp}$
lifts the degeneracy, and the minima moves to $(\pi/2,\pi/2)$ \cite{fren}.
At the same time the energy at $k=(0,0)$ coincides with that at $(\pi,0)$.
It is worth noting that the approximate relation
$E(\pi/2,0)= 0.5 E(\pi/2,\pi/2)$ holds.
An analogous one-hole dispersion relation was obtained in \cite{chuka}
by doing small-U perturbation theory of the one-band Hubbard model.

Consider now the effect of interaction.
 Equation (\ref{A}) was solved numerically by the simplest iteration procedure.
The Green's function at energy $\omega$ on the right hand side of (\ref{A})
was used to calculate the self-energy at the next energy slice,
$\omega + \Delta \omega$. We choose $\Delta \omega$ to be $0.01$ in
units of $t_a + t_b$, and introduce the damping constant $\delta= 0.01$
to achieve numerical stability.
 To check for finite size effects, we used an integration with 17 points
for the 'best fit' (see below), and found
 no practical difference with the 13-points integration.
To check the validity of the
procedure we used it to solve the self-consistent equation for the
$t-J$ model, and found good agreement with existing results \cite{manous}.

 Our 'best fit' to the experimental data is shown on Fig.1. Parameters
of the fit are the following: $t_a=0.19, ~t_b=0.5, ~t_{pp}=0.76,
{}~J=0.125$ in eV. It is quite easy to achieve a reasonable agreement with
experiment along $(0,0)\rightarrow(\pi/2,\pi/2)\rightarrow(\pi,0)$
directions. The main problem is to fit experiment around the $(\pi/2,0)$
point. Our search for the 'best fit' mainly consisted in minimizing
the deviation between Eq.(\ref{A}) and experimental data in the region
of momenta between $(\pi,0)$ and $(\pi/2,0)$. We found that spectrum
$E(k)$ is not affected by small variations of $t_a$ and $t_b$ while
keeping their sum fixed. For example, $t_a=0.07, ~t_b=0.62, ~t_{pp}=0.76$
fits the experimental $E(k)$ as well as the 'best fit'. We found that even
better agreement along $k_y=0$ line can be reached by choosing
unrealistical values of parameters: $t_a=0.31,~t_b=0.94,~t_{pp}=1.37$.
But the bandwidth gets higher for these values (Fig.2).

 The spectral density $A(k,\omega)=\frac{1}{\pi} Im(Tr G(k,\omega))$
for the 'best fit' is shown on Fig.3. Quasiparticle peaks are
observed at all values of $k$. At $k=(0,0)$ the weight of the quasiparticle
pole is smallest, most of the weight is concentrated at the very
strong peak at $\omega=0$ \cite{kaban}.
 To see that lowest peaks of $A(k,\omega)$ indeed represent quasiparticles
we analyzed the $\delta$-dependence of peaks at $k=(\pi/2,\pi/2)$, $(\pi,0)$
 and $(\pi/2,0)$.
In all cases peak fits well by a Lorentzian
$\frac{1}{\pi} \frac{Z\delta}{(\omega - \omega_{peak})^2 + \delta^2}$
\cite{manous}. As expected, the residue $Z(k)$ is strongest at hole's
minima, $Z(\pi/2,\pi/2)\simeq 0.25$, whereas $Z(\pi/2,0)$ is 4 times smaller.
It is even smaller then that at the $(\pi,0)$ point
$(\frac{Z(\pi/2,\pi/2)}{Z(\pi,0)} \simeq 6)$ simply because quasiparticle's
energy at this momentum is higher then that at the $k=(\pi/2,0)$.
It is worth noticing that our value of $Z(\pi/2,\pi/2)$ is quite close
to that in the t-J model \cite{manous}. Smallness of the ratio
$\frac{Z(\pi/2,0)}{Z(\pi/2,\pi/2)}$ indicates that experimental determination
of the hole's energy around $k=(\pi/2,0)$ can be difficult.

 Recently Nazarenko et al. \cite{nazar} reported theoretical analysis
of the same problem. They showed numerically that the spectrum isotropy
around the band minima can be explained within the single-band
$t-J$ model with the small value of intrasublattice hopping $t' \sim J$.
However 'flatness' around $(\pi/2,0)$ was not explained. Also
they (see also \cite{vos}) performed calculations of the $E(k)$
based on the variational wavefunction for the three-band model, and
reached an impressive agreement with experimental results at all
values of $k$. Theoretical reason for such agreement is not clear to
us. The only difference with our analysis is that they also included
$\it{local ~Kondo-like}$ coupling between $Cu$ and $O$ spins
($J_2$ term in notations of Ref.\cite{fren}, or $K_{pd}$ in notations
of \cite{quan-chem}). Notice that such local
Kondo-term has no effect on the spectrum at the mean-field level, because
it is proportional to
$(\vec{S}_m + \vec{S}_{m+2g})\cdot \vec{\sigma}_{m+g}$, which is zero
on the Neel state. On the quantum level this term does affect structure
of Eq.(\ref{A}) significantly, as it couples magnons with both $c$- ~and
$z$-fermions. We have solved Eq.(\ref{A}), reformulated to take into account
$J_2$ coupling, and found no $qualitative$ changes in the $E(k)$.

 To make a connection with existing estimates of the microscopical
parameters based on the density-functional approach we present on Fig.4
results of our calculations with parameters from Hybertsen et al.
\cite{quan-chem}. Clearly, the agreement with experiment is reasonable.

 To summarize, we found that experimental results of Wells et al.\cite{wells}
can be reasonably well described within the three-band model with
values of microscopical parameters in good agreement with existing
estimates. The main result is that direct oxygen-oxygen
hopping $t_{pp}$ is the largest parameter in the problem. This is
necessary for explaining the isotropy of the spectrum arount the minima
at $(\pi/2,\pi/2)$. At the same time our results show the presence
of the quasiparticle pole in the Green's function at $k=(\pi/2,0)$
also, whereas experimentally the possibility of no coherent excitations
at this value of momentum can not be ruled out. Higher
experimental accuracy is necessary to clarify this question.

%\section{acknowledgements}
We thank B. Wells for providing us with experimental data.
We also thank A.Chubukov, D.Frenkel, and K.Musaelian for discussions.
We are grateful to R.Gooding and K.Vos for sending us their
results and useful discussions.

\begin{figure}
\caption{
'Best fit' to the experimentally measured quasiparticle dispersion.}
\end{figure}

\begin{figure}
\caption{Example of fit with 'unrealistic' values of parameters.}
\end{figure}

\begin{figure}
\caption{
Spectral function of hole at different momenta. Parameters correspond
to the 'best fit', Fig.1. $\omega$ is in units of $t_a + t_b$.
 }
\end{figure}

\begin{figure}
\caption{ Quasiparticle dispersion with parameters based on
density-functional calculations, Ref.[14].
}
\end{figure}

\end{document}